\input epsf
\documentstyle[twoside,fleqn,npb,epsfig]{article}
   \newcommand{\beq}{\begin{equation}}
   \newcommand{\eeq}{\end{equation}}
   \def\e{\epsilon}

\newcommand{\AmS}{{\protect\the\textfont2
  A\kern-.1667em\lower.5ex\hbox{M}\kern-.125emS}}

\title{Urgent problems at small $x$}

\author{P V Landshoff\address{DAMTP, Cambridge University \\
        Cambridge CB3 9EW, England \\
        pvl@damtp.cam.ac.uk}%
        \thanks{This research is supported in part by the EU Programme
``Training and Mobility of Researchers", Networks
``Hadronic Physics with High Energy Electromagnetic Probes"
(contract FMRX-CT96-0008) and
``Quantum Chromodynamics and the Deep Structure of
Elementary Particles'' (contract FMRX-CT98-0194),
and by PPARC}}

\begin{document}

\begin{abstract}
Regge theory provides an excellent fit to small-$x$ structure-function data
from $Q^2=0$ right up to the highest available values, but it also
teaches us that conventional approaches to perturbative evolution are
incorrect.
\end{abstract}

\maketitle

\section{INTRODUCTION}

During the 1960's, a lot was learnt about the analytic properties of
scattering amplitudes. Much of this knowledge was incorporated in
Regge theory, but it has been largely forgotten. However, Regge
theory provides\cite{twopom} the best available description of
the structure function data at small $x$, right from $Q^2=0$ up to the very
highest available values. It should not be regarded as a competitor for
perturbative QCD; rather, it is complementary to it, and we need to
learn how to make the two live together. In recent years, a belief has
grown that the spectacular small-$x$ behaviour seen at HERA 
may be associated with the collinear singularity of the DGLAP splitting
function\cite{esw}. However, this belief conflicts with what we know about
the analytic properties of the structure function\cite{cdl}.
\section{REGGE FIT TO SMALL-$x$ DATA}

Regge theory should be valid at any value of $Q^2$, provided only
that $x$ is small enough. 
In its simplest form, it describes the structure function as a sum
of fixed powers of $x$, multiplied by functions of $Q^2$:
\beq
F_2(x,Q^2)\sim\sum _if_i(Q^2)x^{-\e _i}
\label{one}
\eeq
Regge theory tells us little about the coefficient functions
$f_i(Q^2)$, beyond that they are analytic functions of $Q^2$.
Also, we know from QED gauge invariance
that at $Q^2=0$ they vanish at least linearly with $Q^2$.
It may be that the assumption of simple powers of $x$ is
too simple, and it certainly must be corrected at some level,
but it does fit the data extraordinarily well and so there is
no reason to suppose that the correction is numerically significant
at present $x$ values.
We find that three powers are sufficient: two are taken from our 
old fits\cite{sigtot} to hadronic total cross sections
\begin{eqnarray}
\e _1&=0.08 ~~~&\hbox{ soft pomeron exchange} \\
\e _2&=-0.45 ~~~&f,a \hbox{ exchange} 
\end{eqnarray}
The data require the remaining power to be
\beq
\e _0=0.4
\label{three}
\eeq
with an error of about $\pm 10\%$. We call this the ``hard pomeron''.

We have made a fit\cite{twopom} to the data at each available value of $Q^2$
to extract the values of the coefficient functions $f_i(Q^2)$.
The data do not constrain the $f,a$-exchange coefficient function
$f_2(Q^2)$ at all well, but the result for  the hard-pomeron function
$f_0(Q^2)$ and the soft-pomeron function $f_1(Q^2)$ are shown in figure~1.

\begin{figure}[htb]
\epsfxsize=75truemm\epsfbox{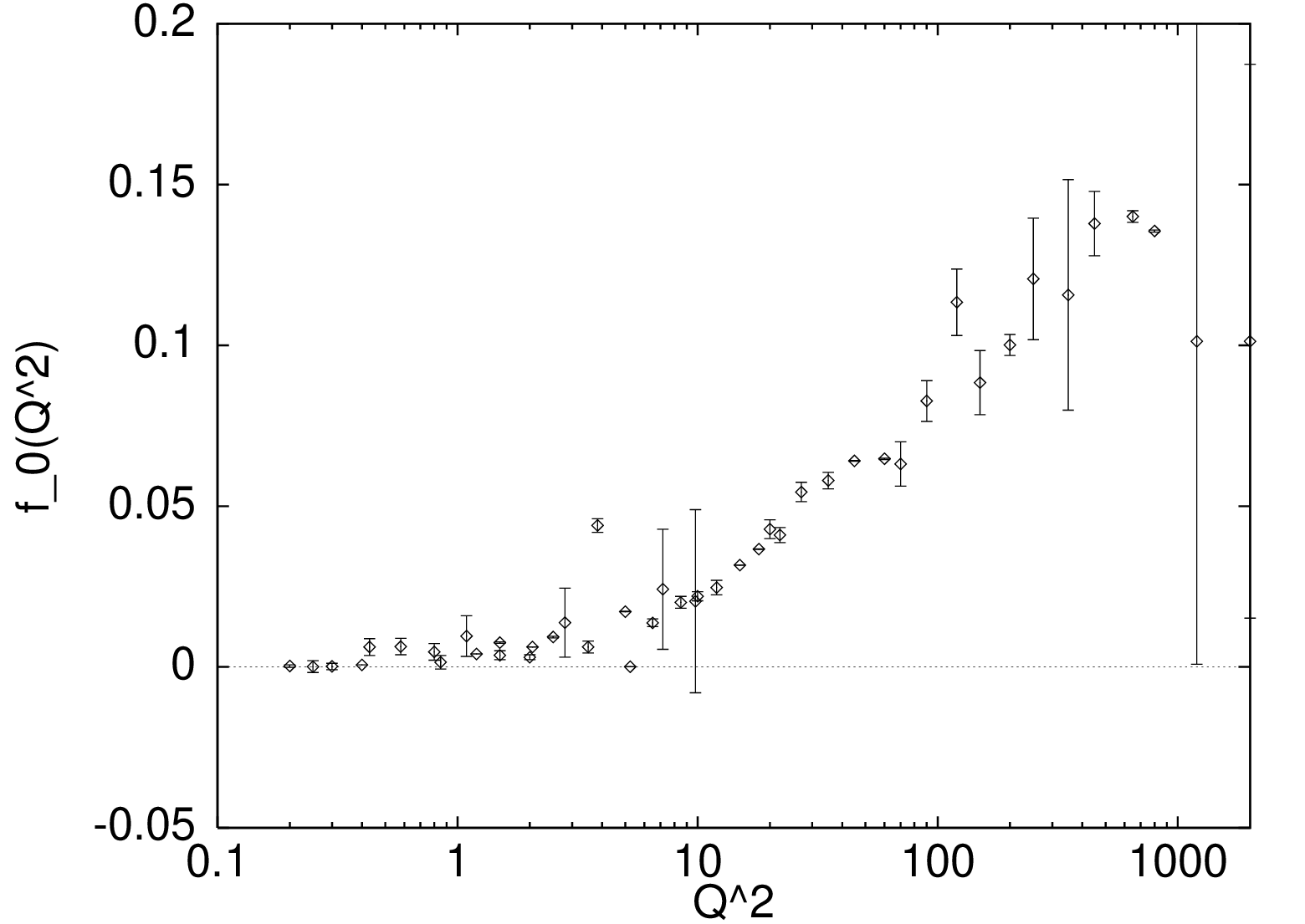} \vskip 10pt
\epsfxsize=75truemm\epsfbox{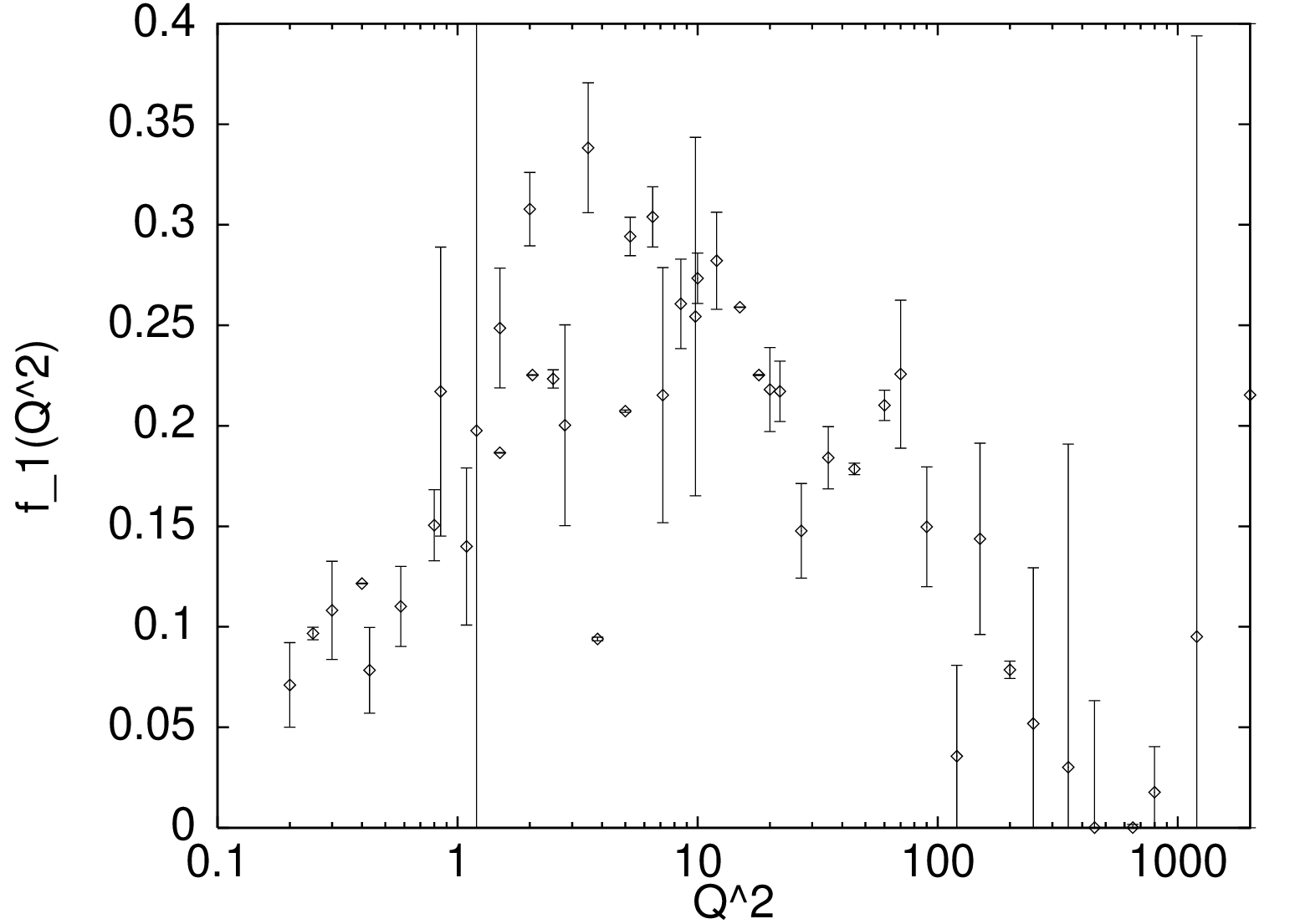}
\caption{The coefficient functions $f_0$ and $f_1$ extracted from
data at each $Q^2$; the error bars are from MINUIT}
\end{figure}

Each vanishes at $Q^2=0$, as it has to. The hard-pomeron coefficient
remains small until about $Q^2=10$ GeV$^2$, after which it rises approximately
logarithmically. This is no surprise. What is surprising is that the
soft-pomeron coefficient, after rising rapidly away from 
$Q^2=0$, reaches a peak at about $Q^2=10$ and then falls again. That is,
soft-pomeron exchange is higher twist. For even quite large values of $Q^2$
this higher-twist component is a major part of the small-$x$ structure
function: see figure 2. This raises serious questions about all
perturbative-QCD fits to structure functions.

\begin{figure}
\epsfxsize=65truemm\epsfbox{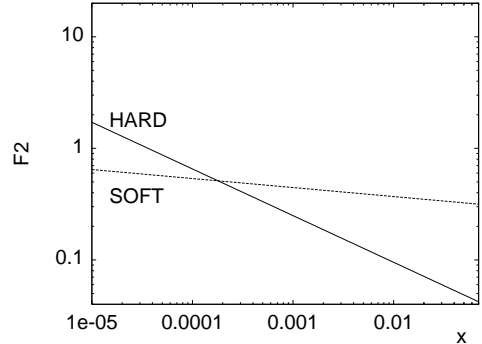}
\caption{Hard and soft contributions to $F_2(x,Q^2)$ at $Q^2=5$}
\end{figure}

\begin{figure}[htb]
\epsfxsize=75truemm\epsfbox{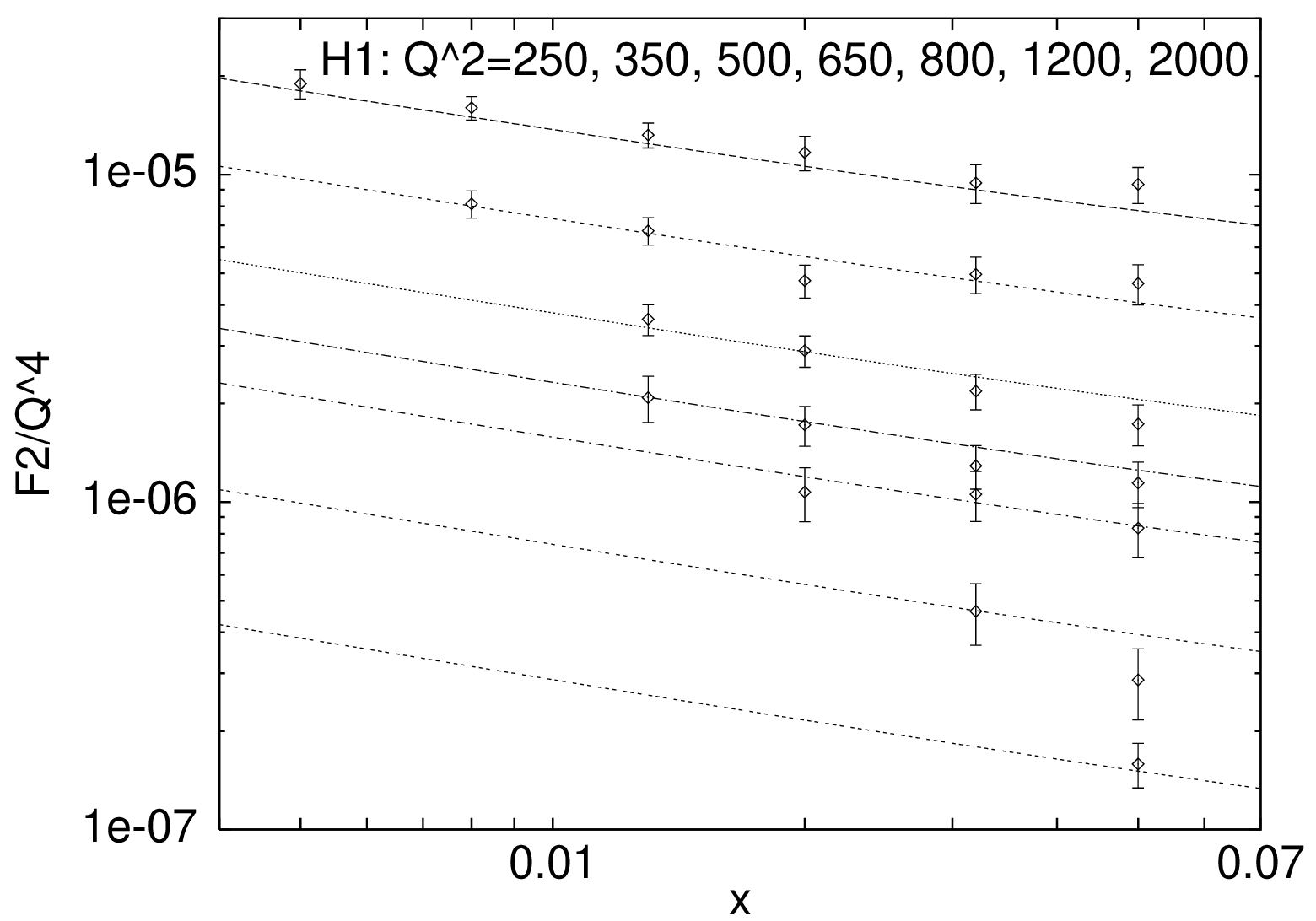} \vskip 10pt
\epsfxsize=75truemm\epsfbox{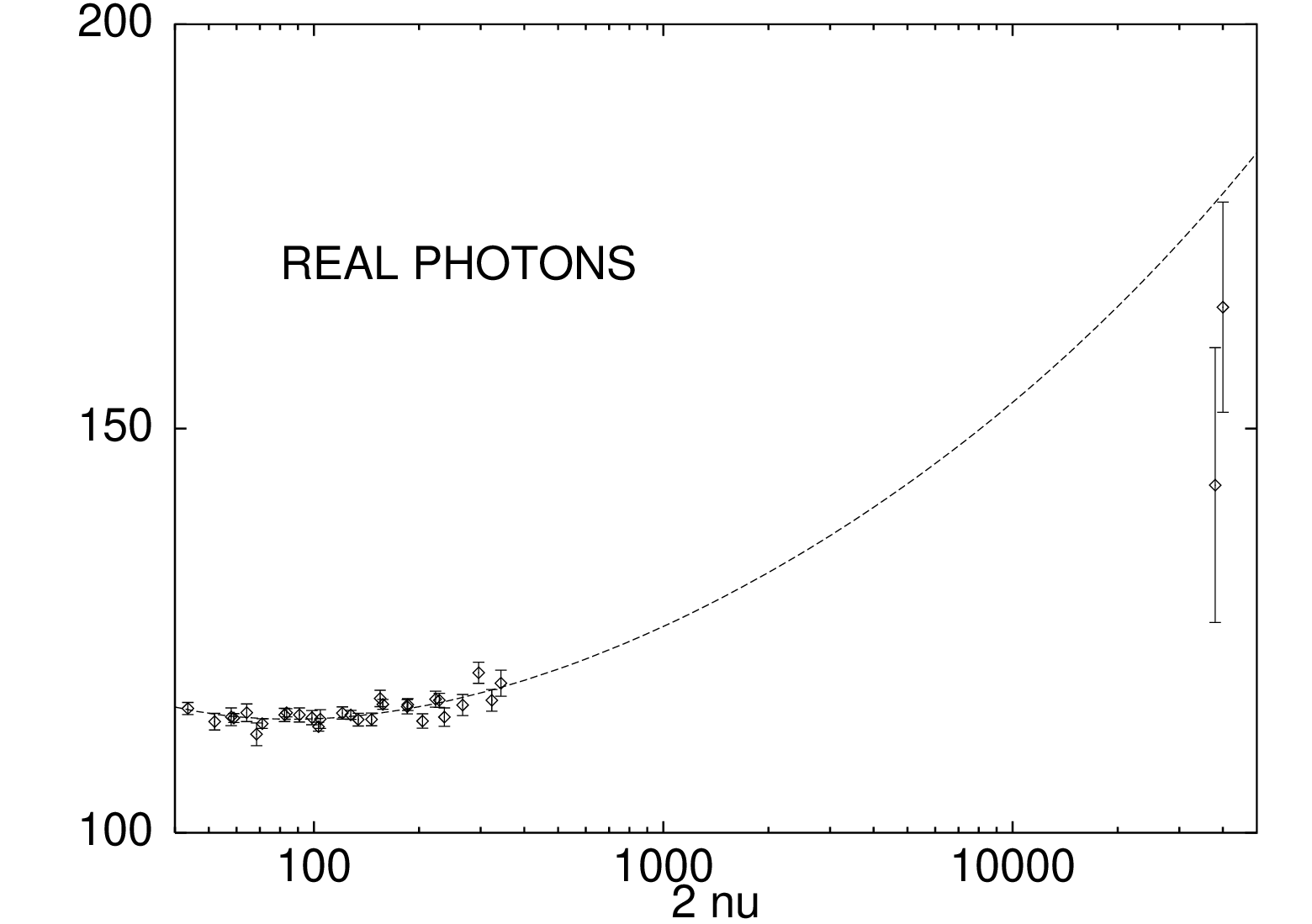}
\caption{Regge fits to the largest-$Q^2$ data and the real-photon data}
\end{figure}

The three-term form (1) gives an excellent fit to $F_2(x,Q^2)$
for $x<0.07$ and $0\leq Q^2 \leq 2000$. With 8 free parameters,
including $\e _0$, one can achieve a $\chi ^2$ per data point 
well below 1.0, so the exact values of the parameters are not
completely determined by these data points.  Figure 3 shows how
such a fit compares with the largest-$Q^2$ data  and the real-photon data.
A combination of the hard and the soft pomerons also describes well\cite{twopom}
the data for the process $\gamma p\to\psi p$.

\section{PERTURBATIVE EVOLUTION}

\def\P{{\bf P}}
\def\f{{\bf f}}
\def\u{{\bf u}}
\def\pd{\partial}
If we Mellin transform with respect to $x$, the DGLAP equation\cite{esw}
reads
\beq
{\pd\over\pd\log Q^2}\u (N,Q^2)= \P(N,Q^2)\u (N,Q^2)
\label{dglap}
\eeq
where  $\u$ is a two-component object whose elements are the singlet
quark distribution and the gluon distribution, while $\P$ is the splitting
matrix. A power
contribution
\beq
f(Q^2)x^{-\e}
\eeq
to $F_2(x,Q^2)$ corresponds to a pole
\beq
{\f(Q^2)\over N-\e}
\eeq
in $\u(N,Q^2)$.  Inserting such a pole into each side of (\ref{dglap})
gives a differential equation for $f(Q^2)$. If we use the lowest-order
approximation to the splitting matrix $\P$ the solution to this equation
is that, for large $Q^2$, $f(Q^2)$ behaves\cite{ynd} as a power of 
$\log Q^2$.

However, there is a serious problem with this lowest-order approximation.
It gives $\P(N,Q^2)$ a pole at $N=0$, and it is this pole
that largely determines the magnitude of the power of $\log Q^2$. 
Conventional
perturbative-QCD fits to the data also rely on this pole to explain the
rapid rise of $F_2$ at small $x$.  But we know that in fact such a pole
cannot be present: higher-order corrections must resum it away.
We know this because at small $Q^2$ $\u (N,Q^2)$ does not have a
singularity at $N=0$: rather, its singularities in the complex $N$-plane
are the
standard singularities
of Regge theory --- the soft pomeron, the mesons, and possibly also a
hard pomeron. It is also supposed to be analytic in $Q^2$, so a singularity
at $N=0$ cannot suddenly appear when we continue from small $Q^2$ up
to beyond the values of $Q^2$ at which the DGLAP equation (\ref{dglap})
begins to be valid. Thus $\P(N,Q^2)$ cannot have a pole
at $N=0$, nor indeed at any other value of $N$.

At small $N$, the $gg$ element of the splitting matrix is found by
solving the equation\cite{jaros}
\beq
\chi (P_{gg}(N,Q^2),Q^2)=N
\eeq
where $\chi(\omega, Q^2)$ is the Lipatov characteristic function. In lowest
order,
\beq
\pi\chi(\omega, Q^2)=3\alpha_S(Q^2)\,[2\psi(1)-\psi (\omega)-\psi (1-\omega)]
\eeq
If one uses this approximation to $\chi(\omega, Q^2)$ one indeed finds that
$P_{gg}(N,Q^2)$ is nonsingular at $N=0$, even though the terms of its expansion
in powers of $\alpha _S$ are each singular at $N=0$. Compare the expansion
of the function
\beq
p(N,Q^2)=N-\sqrt{N^2-\alpha _S(Q^2)}
\eeq
whose expansion is
\beq
p(N,Q^2)={\alpha _S(Q^2)\over 2N}+{\alpha ^2 _S(Q^2)\over 8N^3}+\dots
\eeq
but which is evidently finite at $N=0$. Near $N=0$ the expansion
parameter $\alpha_S(Q^2)/N$ is so large that the expansion is illegal.

We know that $P_{gg}(N,Q^2)$ is finite at $N=0$, but we do not
know how large it is, because the lowest-order approximation (9) to
$\chi(\omega, Q^2)$ is apparently not a good one: the next-to-leading
order correction is huge\cite{lipatov}

More work is needed!

\end{document}